\documentclass[twocolumn, superscriptaddress]{revtex4-1}
\usepackage{amsmath}
\usepackage{graphicx}
\usepackage{natmove}
\usepackage{dcolumn}
\usepackage{hyperref}
\usepackage{afterpage} 
\usepackage{array}
\usepackage{soul,color}

\begin{document}

\title{Uncertainty Quantification for Materials Properties in Density Functional Theory with k-Point Density}

\author{Joshua J. Gabriel} 
\affiliation{Department of Materials Science and Engineering, University of Florida, Gainesville, Florida 32611}

\author{Faical Yannick C. Congo} 
\affiliation{Material Measurement Laboratory, National Institute of Standards and Technology, 100 Bureau Drive, Gaithersburg, Maryland 20899}
\affiliation{ISIMA/LIMOS UMR CNRS 6158}
\affiliation{Blaise Pascal University - Campus Universitaire des Cezeaux, 2, Rue de la Chebarde, TSA 60125 - CS 60026, 63178 Aubière CEDEX FRANCE}

\author{Alexander Sinnott}
\affiliation{Department of Materials Science and Engineering, University of Florida, Gainesville, Florida 32611}

\author{Kiran Mathew} 
\affiliation{Department of Materials Science and Engineering, Cornell University, Ithaca, New York}

\author{Thomas C. Allison} 
\affiliation{Material Measurement Laboratory, National Institute of Standards and Technology, 100 Bureau Drive, Gaithersburg, Maryland 20899} 

\author{Francesca Tavazza} 
\affiliation{Material Measurement Laboratory, National Institute of Standards and Technology, 100 Bureau Drive, Gaithersburg, Maryland 20899} 

\author{Richard G. Hennig}
\affiliation{Department of Materials Science and Engineering, University of Florida, Gainesville, Florida 32611} 
%\email{rhennig@ufl.edu}

\begin{abstract}
Many computational databases emerged over the last five years that report material properties calculated with density functional theory. The properties in these databases are commonly calculated to a precision that is set by choice of the basis set and the $k$-point density for the Brillouin zone integration. We determine how the precision of properties obtained from the Birch equation of state for 29 transition metals and aluminum in the three common structures -- fcc, bcc, and hcp -- correlate with the $k$-point density and the precision of the energy.  We show that the precision of the equilibrium volume, bulk modulus, and the pressure derivative of the bulk modulus correlate comparably well with the $k$-point density and the precision of the energy, following an approximate power law.  We recommend the $k$-point density as the convergence parameter because it is computationally efficient, easy to use as a direct input parameter, and correlates with property precision at least as well as the energy precision. We predict that common $k$-point density choices in high throughput DFT databases result in precision for the volume of 0.1\%, the bulk modulus of 1\%, and the pressure derivative of 10\%.
\end{abstract}

%\begin{keyword} 
%Uncertainty quantification; density functional theory; Brillouin zone integration
%\end{keyword}

\maketitle

\section{Introduction}

Computational databases of material structures and properties are an essential data source for materials discovery and design \cite{Meredig2017, Jain2016_Review,2016_Citrine1, Meredig_2015, Yin1_2015,Ceder2012Batteries} as they allow high throughput screening for specific properties.  These databases also provide input data for larger scale simulations and experimental work, validation data, and property values to aid the interpretation of experimental results.  Inspired by the Materials Genome Initiative, recently many repositories emerged that provide structures and calculated properties for materials, ranging from bulk metals and inorganic compounds, {\it e.g.}, the Materials Project \cite{Jain2013}, Aflowlib \cite{curtarolo2012aflowlib}, OQMD \cite{OQMD1}, JARVIS \cite{choudhary2018elastic}, to polymers \cite{Rampi2016}, and 2D materials \cite{Ashton2017a} . All these databases have in common that they contain structures and properties of both experimental and theoretically predicted materials and that the materials properties are obtained from density-functional theory (DFT) calculations.

Building on these databases and fulfilling the vision of the Materials Genome Initiative, there is a need to quantify the epistemic uncertainties in computed material properties that are derived from density functional calculations \cite{pernot2015prediction, chernatynskiy2013uncertainty}. The uncertainties can be classified into the precision of the DFT calculation and the accuracy of the underlying DFT method. Knowledge of the accuracy for property predictions is important for the use of databases in materials selection and design. Knowledge of the precision of the computed property can help identify reliable trends across material families and ensure reliable property predictions. We note that the precision is different from the relative precision related to the reproducibility of property predictions across different DFT codes \cite{lejaeghere2016reproducibility}. This relative precision uses the $\Delta$ gauge to compare the energy vs. volume curves between any two codes.  The precision quantifies the uncertainty in the computed property value due to the choice of convergence parameters, such as the basis set and $k$-point mesh density for the Brillouin zone integration required for crystalline materials. The accuracy of DFT calculations is controlled by the choice of exchange-correlation functional and pseudopotential. The accuracy is calculated with respect to experiment or high-quality calculations \cite{Draxl2018, Parker2010, OnemevAccuracy1, Tipton_PseudoPotentials_PRB2014}.

While several studies investigate the accuracy of material properties predicted by DFT due to the choice of exchange-correlation functional and pseudopotential \cite{Parker2010, OnemevAccuracy1, Tipton_PseudoPotentials_PRB2014, Cohen_Accuracy2006, TruhlarAccuracy_2008, Perdew_Accuracy, PerdewAccuracy_2009, Shang2010, Janthon2013, Tran2016}, little is known about the precision of the computed properties. Some studies have shown that DFT can achieve $\mu$Ha precision in the energies, through optimization of the basis set or the $k$-point density \cite{Blum2017, Draxl2018, Morgan2018}. Error bars have been assigned to DFT computed properties based on systematic trends in the accuracy of material properties \cite{Lejaeghere2014}, but little is known about error bars due to the user inputs into a DFT calculation alone. Additionally, there is a growing need for guidance on convergence rules that translate into user inputs, much like advice is available for the choice of interatomic potentials for molecular dynamics simulations \cite{becker2013considerations}.

Increasing the precision, {\it i.e.}, minimizing the precision error, comes with an increase in computational cost. High throughput calculations of computational materials databases trade precision in the calculated materials properties for the number of entries in the database, given a computational budget. Convergence parameters, like the basis set size and the $k$-point density, are commonly determined from calculations for a diverse subset of materials. High-throughput approaches then apply these parameter choices across a large number of materials and do not test or change them for individual materials. This approach simplifies the workflow, reduces computational cost, and results in partial error cancellation for incomplete basis sets. However, with this choice of convergence parameters it is difficult to quantify the precision error for the computed property values for each material.

An alternative, frequently applied convergence strategy is to increase the computational parameters such as the basis set size and $k$-point density until the energy changes by less than a predefined value, {\it e.g}, 1~meV/atom, for successive choices of the parameter. This convergence criterion is empirically motivated by observation of energy differences of structural phase transformation in materials and few other materials properties \cite{OnemevAccuracy1, OnemevAccuracy2}. However, this energy convergence criterion may not guarantee a desired precision in the prediction of other material properties derived from the energy. For example, for calculations of elastic constants de Jong {\it et al.} considered the residual stresses and the forces on the atoms as an additional criterion to the energy convergence \cite{MP_properties_deJong2015}.

In this paper, we investigate the precision error of the energy, structural and elastic properties derived from an equation of state, namely, the cohesive energy, $E_0$, equilibrium volume, $V_0$, bulk modulus, $B$, and its pressure derivative, $B'$. Sec.~\ref{sec:Method} describes our computational workflow to obtain the properties as a function of $k$-point density and their extrapolated values. We use the extrapolated value of each property as the reference value to calculate the precision error. Sec.~\ref{sec:Results} analyzes the convergence of the properties and precision with $k$-point density. In Sec.~\ref{Extrapolates} we show that the uncertainty of the extrapolated properties is similar for different choices of exchange-correlation functional and DFT code. In Secs.~\ref{Kpoints} and ~\ref{Energy} we show that the precision of these derived properties correlates with both the $k$-point density and the precision of the energy following a power law. We introduce a Pareto optimality method to determine the minimum $k$-point density choice, or the maximum precision in the energy, which ensures a desired precision for each property. Finally, we predict the expected precision of these derived properties for common choices of $k$-point density in high-throughput DFT databases. We also predict the expected precision for similar choices of the energy convergence.  

\begin{figure}
\includegraphics[width=\columnwidth]{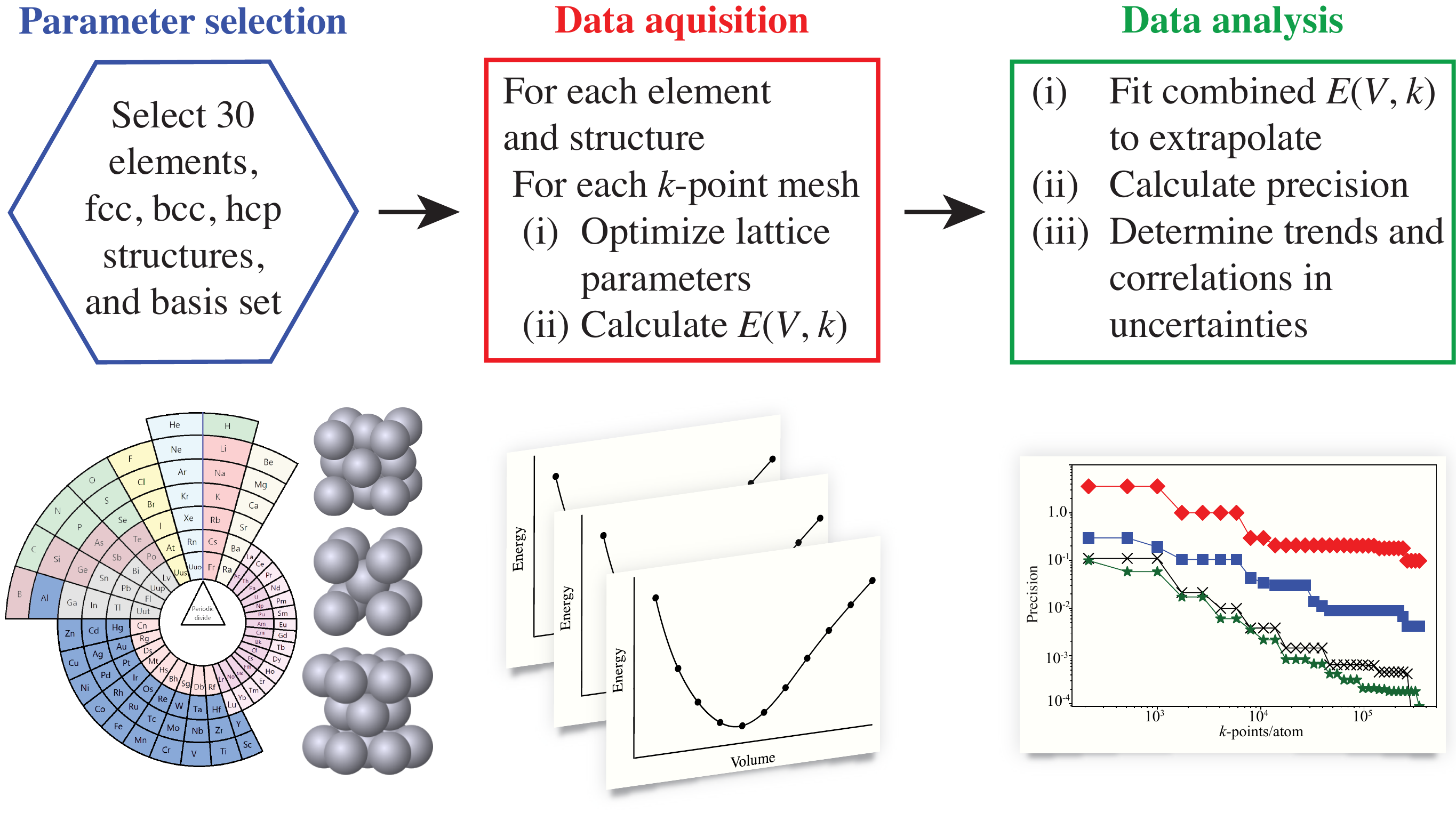}
\caption{\label{fig:workflow} Our data acquisition and uncertainty quantification workflow (left to right): (1) We select 30 elements and determine their equation of state for the fcc, bcc, and hcp structures using DFT calculations for $k$-point densities ranging from about 10 to $10^5$ pra. (2) We extrapolate the $E(V,k)$ data with Eq.~\eqref{eq:Combined} to infinite $k$-point density. We use the extrapolated material properties to calculate the precision of each property using Eq.~\eqref{eq:precision_E} and Eq.~\eqref{eq:precision_P}. (3) We analyze the convergence behavior for the precision with Eq.~\eqref{eq:PowerKpoints} and Eq.~\eqref{eq:PowerEnergy}, and determine a Pareto optimal choice of the $k$-point density for each property using Eq.~\eqref{eq:ParetoChoiceK} and choice of energy convergence criteria using Eq.~\eqref{eq:ParetoChoiceE}.}
\end{figure}

\section{Computational Method}
\label{sec:Method}

Fig.~\ref{fig:workflow} summarizes our computational workflow for the data acquisition and uncertainty quantification. We apply our data approach to the fcc, bcc and hcp structures of aluminium and the 3d, 4d, and 5d transition metal elements, leaving out lanthanum. For these 90 materials, we calculate the energy vs.\ volume curves with DFT for different $k$-point densities. We describe the DFT calculations in detail in Sec.~\ref{DFT}. We fit the Birch equation of state for the different $k$-point density choices, as described in Sec.~\ref{Property}. In Sec.~\ref{Numerical_Precision}, we outline the extrapolation of the materials properties to infinite $k$-point density by incorporating an exponential function for the convergence of the parameters with $k$-point density into the Birch equation of state. From the extrapolated values, we calculate the precision of the properties. 

\subsection{Density functional theory calculations} \label{DFT}

In our study, we chose two different density functional codes, the plane-wave basis package VASP \cite{VASP1, VASP2}, and the local atomic basis package DMol3 \cite{delley2000molecules}. We select the $k$-point mesh density to be the controlling approximation for our comparison of precision, which can be set to equivalent values in both codes. We choose the tetrahedron method \cite{TetraBloechl} and Monkhorst Pack $\Gamma$-centered meshes \cite{Monkhorst} for the Brillouin integration, which is also common to both codes. We select uniform $\Gamma$-centered $k$-point meshes with densities ranging from 10 to $10^5$ $k$-points per reciprocal atom (pra). $k$-points per reciprocal atom (pra) is defined as the total number of $k$-points divided by the reciprocal of the number of atoms. We note there are other choices to quantify the $k$-point density such as the linear $k$-point density per \AA\ and the volume $k$-point density per cubic \AA\ of the reciprocal lattice volume, as well as other $k$-point mesh choices that can improve efficiency of convergence \cite{Mueller2016,Morgan2018}. We choose the Monkhorst Pack $\Gamma$-centered meshes and the $k$-point density pra as the unit of comparison because this is a common choice among high-throughput DFT databases \cite{Jain2013, Ashton2017a}, with the option to convert to the other units mentioned. We include the $\Gamma$ point in our $k$-point meshes because DMol3 requires an odd mesh, which contains the $\Gamma$ point. For VASP we chose an even-numbered mesh centered around the $\Gamma$ point, which is a  choice made in many high-throughput DFT databases. Henceforth, we refer to the $k$-points pra as the $k$-point density.

We select the Perdew-Burke-Ernzerhof (PBE) \cite{perdew1996generalized} generalized-gradient approximation (GGA) for the exchange-correlation functional because it is common to both codes and one of the most widely used functionals. We also perform calculations with the local-density approximation (LDA), which uses the Perdew-Zunger parameterization \cite{PZ81} in VASP and the Perdew-Wang parameterization in DMol3 \cite{PW92}.

For the basis sets, we perform the DMol3 calculations with the largest basis set available, the double-$\zeta$ plus polarization function basis set and a real space cutoffs of 6~\AA. For VASP, we perform the calculations with a fixed plane-wave energy cutoff of 550~eV, similar to the value used in the Materials Project database of 520~eV \cite{Jain2013}. For the pseudopotential approximation, we employ norm-conserving semilocal pseudopotentials in the DMol3 calculations \cite{Delley2002} and the projector augmented wave method in the VASP calculations \cite{PAW}.

\subsection{Property estimation}
\label{Property}

To estimate the properties at the different $k$-point densities, we adopt a two-step data generation workflow, where we first calculate the equilibrium volume for each material and $k$-point density and second calculate the energy, $E$, as a function of volume, $V$, for 11 equally spaced data points that bracket the calculated equilibrium volume and span a range of $\pm5\%$.

To determine the cohesive energy, $E_0$, the equilibrium volume, $V_0$, the bulk modulus, $B$, and its pressure derivative, $B'$, for each material, we fit the Birch equation of state \cite{Birch1947} to the $E(V)$ data:
\begin{equation} \label{eq:Birch}
  \begin{split}
    E(V) = E_{0} + \frac{9}{8}BV_0 {\left ( { \left( \frac{V_{0}}{V} \right)}^\frac{2}{3} - 1 \right)}^2 + \\ + \frac{9}{16}BV_{0} \left(B' - 4 \right) {\left ( \left( \frac{V0}{V} \right)^{\frac{2}{3}} - 1 \right)}^3.
  \end{split}
\end{equation}
We perform these energy volume calculations on the 90 metals in VASP with the PBE functional and on subsets of 40 in DMol3 with the PBE functional and smaller subsets of 15 in VASP and DMol3 with the LDA functionals. We found Hg to be unstable in the hcp structure, from its energy vs.\ volume plot and exclude the hcp phase of Hg from our dataset. To perform the approximately 50,000 DFT calculations, we assembled a high-throughput workflow using the MPInterfaces framework \cite{mathew2016mpinterfaces}, which automates the generation of the energy volume data. For the cohesive energy, $E_0$, we use the energy of the isolated atom as reference \cite{Lejaeghere2014}.

\begin{figure*}
\includegraphics[width=0.85\textwidth]{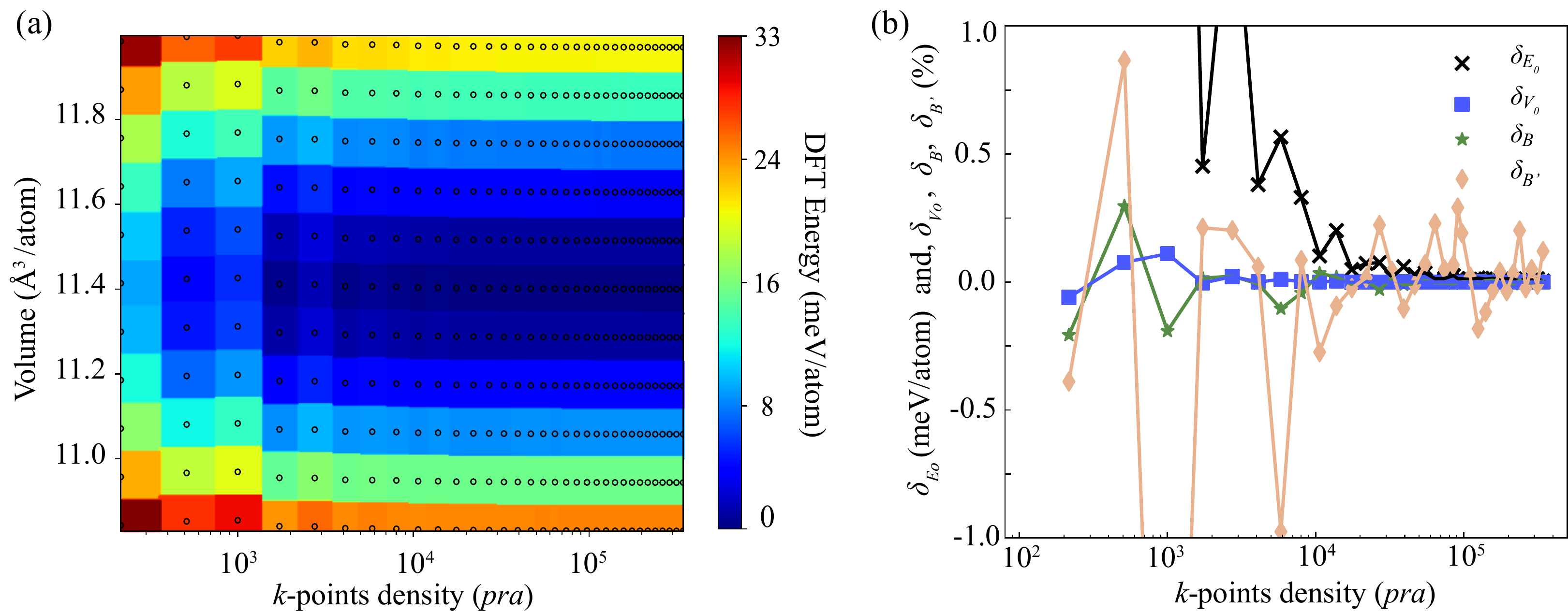}
\caption{\label{fig:method} Calculation of the precision of properties of bcc Cr calculated with VASP in the PBE functional: a contour interpolation of the raw $E(V,k)$ data are shown in (a). From Eq.~\eqref{eq:Birch} for each $k$-point density and the extrapolated values from Eq.~\eqref{eq:Combined}, (b) shows the the precisions in meV/atom for $\delta_{E_0(k)}$ (blue) and in percent error for $\delta_{V_0(k)}$ (orange), $\delta_{B(k)}$ (green), and $\delta_{B'(k)}$ (black), as defined by Eq.~\eqref{eq:precision_E} and Eq.~\eqref{eq:precision_P} as a function of the $k$-point density.} 
%The error bar on the precision is calculated from the fit errors from the equation of state, $\sigma_P$, and the error on the extrapolate, $\sigma_{P_\infty}$. We observe that $\sigma_P$ decreases with increased $k$-point density and is small in comparison to the precision error, $\delta_{P(k)}$.} 
\label{fig:Overview}
\end{figure*}

\subsection{Calculation of precision}
\label{Numerical_Precision}

To calculate the precision, we first choose a reference value. We use the extrapolate of the estimated properties in the limit of infinite $k$-point density as the reference value. For the extrapolation, we tried Pad\'{e}, power-law, and exponential functions. We extrapolate over a variable number of points ranging from the last 5 to 25 $k$-point densities and choose extrapolations that provide a decaying behavior for the precision. We find that the exponential decay function gives a consistent behavior for the decay of the precision with the $k$-point density for all metals considered. By consistent behavior, we mean that the extrapolates obtained using the exponential decay function, gives a similar decaying power law behavior for the absolute value of the precision error as a function of the k-points density.

To minimize the number of fits performed and to directly obtain the error bars on the extrapolated values from the DFT energy vs.\ volume data, we modify the Birch equation of state to account for the $k$-point convergence. We replace the property coefficients, $P$, in the equation of state, Eq.~\eqref{eq:Birch}, with the exponential decay function, 
\begin{align}
\label{eq:Fit}
P(k) &= a + b\exp(-ck) \notag \\
P_\infty &= \lim_{k\rightarrow \infty} P(k) = a,
\end{align}
where the coefficient $a$ provides the extrapolated property value for the limit of infinite $k$-point density. The modified Birch equation of state with each property coefficient replaced with the exponential decay function, $P(k)$, provides now the DFT energy, $E$, as a function of volume, $V$, and the $k$-point density, $k$:
\begin{equation} \label{eq:Combined}
\begin{split}
E(V,k) = E_{0}(k) + \frac{9}{8}B(k)V_{0}(k) {\left ( { \left( \frac{V_{0}(k)}{V} \right)} ^ { \left (\frac{2}{3} \right)} - 1 \right)}^{2} + \\ \frac{9}{16}B(k)V_{0}(k) \left(B^{'}(k) - 4 \right) {\left ( \left( \frac{V_{0}(k)}{V} \right)^{\left (\frac{2}{3} \right)} - 1 \right)}^{3}.
\end{split}
\end{equation}
We fit this equation using a weighted non-linear least-square regression with the Levenberg-Marquardt method \cite{marquardt1963algorithm} as implemented in the R statistics package \cite{Rpackage}.

The fit of the extrapolation function given by Eq.~\eqref{eq:Combined} provides the reference values for each materials property, $P_\infty = a$, from Eq.~\eqref{eq:Fit}. We now estimate the precision for each property, $P$. For the cohesive energy, we calculate the precision $\delta_{E}(k)$, as a simple difference: 
\begin{equation}
  \label{eq:precision_E}
  \delta_{E}(k) = E(k) - E_\infty,
\end{equation}
where $E(k)$ and $E_\infty$ are the values of the cohesive energy at each $k$-point density and the extrapolated value, respectively. For the other properties, {\it i.e.}, $V_0$, $B$, and $B'$ we calculate the precision, $\delta_{P}(k)$ as a percentage error:
\begin{equation}
  \label{eq:precision_P} 
  \delta_{P}(k) = \frac{P(k) - P_\infty}{P_\infty} \cdot 100.
\end{equation}
Fig.~\ref{fig:Overview} illustrates the workflow to obtain the precision of the properties as a function of $k$-point density for bcc Cr.

\section{Results}
\label{sec:Results}

We analyze how the properties and their precisions converge with $k$-point density in three steps. First, Sec.~\ref{Extrapolates} describes the extrapolated material properties, highlighting a minimum error bar, $\sigma_{P_\infty}$, that can be assigned from DFT calculations to each property, $P$. Next, Sec.~\ref{Kpoints} describes the correlation of the precisions with the $k$-point density, $\delta_{P}(k)$. We introduce a Pareto optimal choice of the $k$-point density as a method to determine the minimum $k$-point density for a given required precision. We present the distribution of the precisions that can be expected for common choices of the $k$-point density in high-throughput DFT databases. Finally, Sec.~\ref{Energy} describes the correlation of the  precisions of the properties $V_0$, $B$, and $B'$ with the  precision of the cohesive energy, $\delta_{P}(\delta_{E_0})$. Similar to Sec.~\ref{Kpoints}, we introduce the Pareto optimal choice of the energy precision and the expected  precisions based on choices of 10, 1, and 0.1~meV/atom for the  precision of the cohesive energy.

\subsection{Extrapolated Material Properties}
\label{Extrapolates}
 
The extrapolation of the equation of state using to infinite $k$-point density using Eq.~\eqref{eq:Combined} provides the extrapolated material properties, $E_\infty$, $V_\infty$, $B_\infty$ and $B'_\infty$. In addition, the extrapolation gives the standard error of the coefficients, i.e., the standard error on the respective extrapolated material properties, $\sigma_{E_\infty}$, $\sigma_{V_\infty}$, $\sigma_{B_\infty}$, and $\sigma_{B'_\infty}$ for each metal. This standard error on an extrapolated material property is the minimum error bar on an extrapolated material property. The distribution of this error bar over all the metals, for each property, provides insight into the expected precision that can be obtained by DFT calculations for these properties. Tab.~\ref{tab:extrapolation} shows the 90$^\mathrm{th}$ percentile of the distribution of these error bars across all the materials in this study for the PBE and LDA functionals in VASP and DMol3. We observe that the distribution of error bars of the properties do not strongly depend on the choice of DFT code or exchange-correlation functional. We emphasize that this error bar is based on the extrapolation over the chosen $k$-point densities for the DFT calculation alone and not relative to any experimental reference values.

\begin{table}[b]
\caption{\label{tab:extrapolation} The 90$^\mathrm{th}$ percentile of the distribution of the standard error of the extrapolated cohesive energy $\sigma_{E_\infty}$ in $\mu$eV/atom, and, equilibrium volume $\sigma_{V_\infty}$, bulk modulus $\sigma_{B_\infty}$, and pressure derivative $\sigma_{B'_\infty}$, in percent units of the respective extrapolated values.}
\begin{ruledtabular}
\begin{tabular}{ldddd}
System & \multicolumn{1}{c}{$E_0$ ($\mu$eV/atom)}   & \multicolumn{1}{c}{$V_0$(\%)} & \multicolumn{1}{c}{$B$ (\%)} & \multicolumn{1}{c}{$B'$ (\%)} \\
\colrule
\multicolumn{5}{c}{ ----- PBE ----- } \\
VASP  &  70 & 0.002 & 0.001 & 1.8  \\
DMol3 &  42 & 0.004 & 0.002 & 1.4 \\
\colrule
\multicolumn{5}{c}{ ----- LDA ----- } \\
VASP  &  17 & 0.003 & 0.002 & 1.8 \\ 
DMol3 &  31 & 0.002 & 0.002 & 1.2 \\
\end{tabular}
\end{ruledtabular}
\end{table}

To validate our DFT calculations and workflow, we compare the extrapolated properties with data by Lejaeghere {\it et al.} \cite{lejaeghere2016reproducibility}. We find that the agreement between the extrapolated properties calculated with VASP and DMol3 is comparable to the agreement between plane-wave and local atomic basis set codes. We do not consider spin-polarization in our study and hence obtain different results for Cr, Mn, Fe, and Co.

We also compare the extrapolated properties obtained for the two different functionals, LDA and PBE with both VASP and DMol3. We observe that LDA overbinds compared to GGA-PBE, {\it i.e.}, LDA predicts smaller lattice parameters and higher bulk moduli \cite{OnemevAccuracy1}. For each code, we observe that the mean deviation of the extrapolated properties between LDA and GGA is negative for $V_0$ and $B'$ and positive for $B$. For VASP, the mean deviation for $V_0$ is $-1.0$~\AA$^3$/atom, for $B$ is 47~GPa, and for $B'$ is $-0.1$.

\begin{figure*}
  \includegraphics[width=0.65\textwidth]{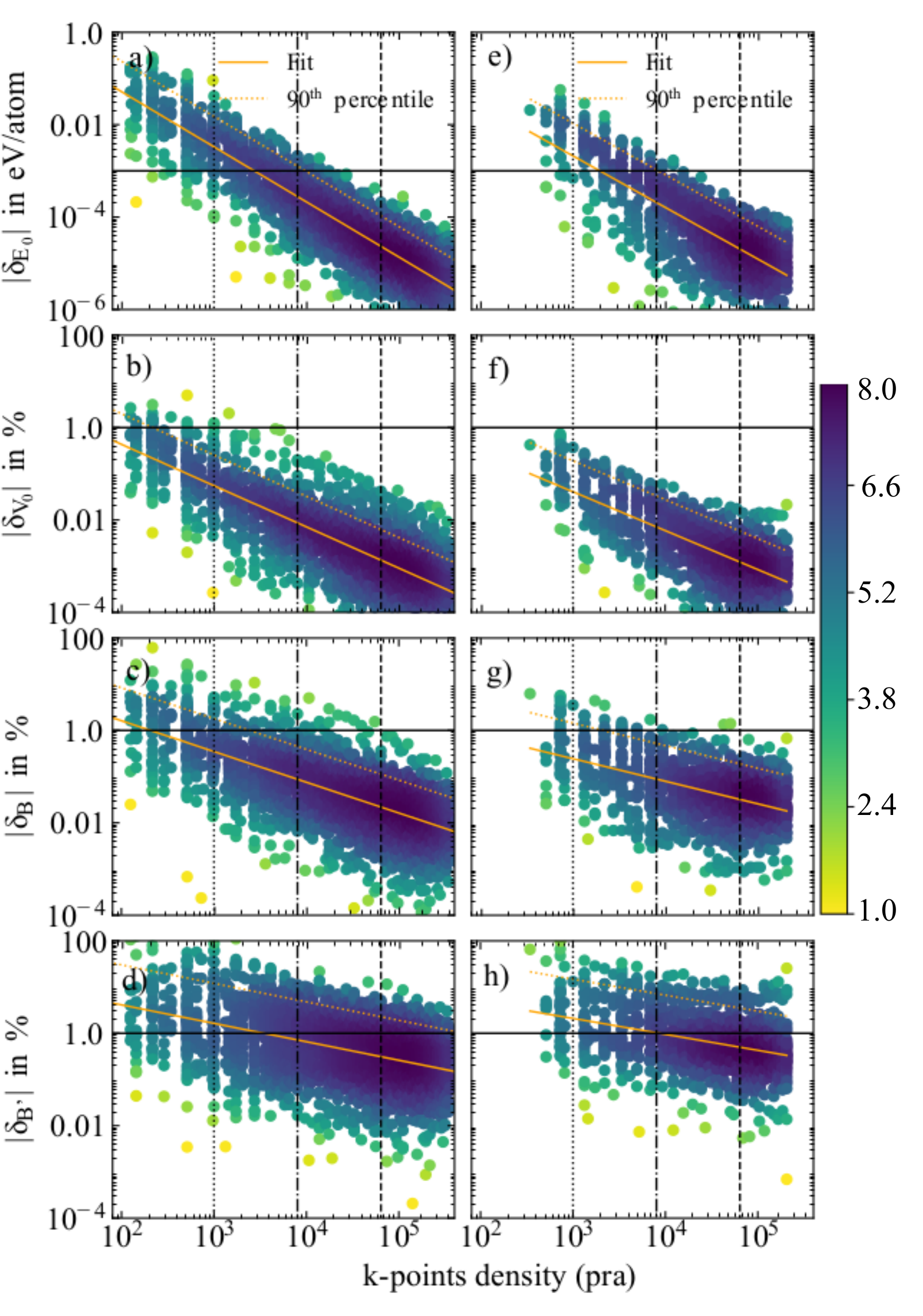}
  \caption{ \label{fig:Complete_K_Convergence} Correlation of precision error with $k$-point density:
    The precision of $E_{0}$ as a function of the $k$-point density is shown as dots shaded from yellow to blue, calculated with the PBE functional using (a) VASP for 90 metals and (b) DMol3 for a subset of 40 metals. Similarly, the precision of $V_0$ is shown in (b) and (f), of $B$ in (c) and (g), and of $B'$ in (d) and (h).  The shading of the dots corresponds to the Gaussian kernel density estimate of the density of data points, yellow being one-eight the density as the dark blue dots. The solid orange lines show the fit of Eq.~\eqref{eq:PowerKpoints}. The 90$^\mathrm{th}$ percentile (dotted orange line) is a trend of the same slope as Eq.~\eqref{eq:PowerKpoints} drawn such that 90\% of the data points fall below the line. The vertical dotted lines indicate the three different $k$-point density choices of 1,000, 8,000 and 64,000 $k$-points pra. For a $k$-point density of 8,000 pra $\delta_{E_0}(k) < 0.001$ eV/atom, $\delta_{V_0}(k)<0.1$\%, $\delta_{B}(k)<1$\%, and $\delta_{B'}(k)<10$\% for 90\% of the metals.}
\end{figure*}

\begin{figure*}
  \includegraphics[width=0.85\textwidth]{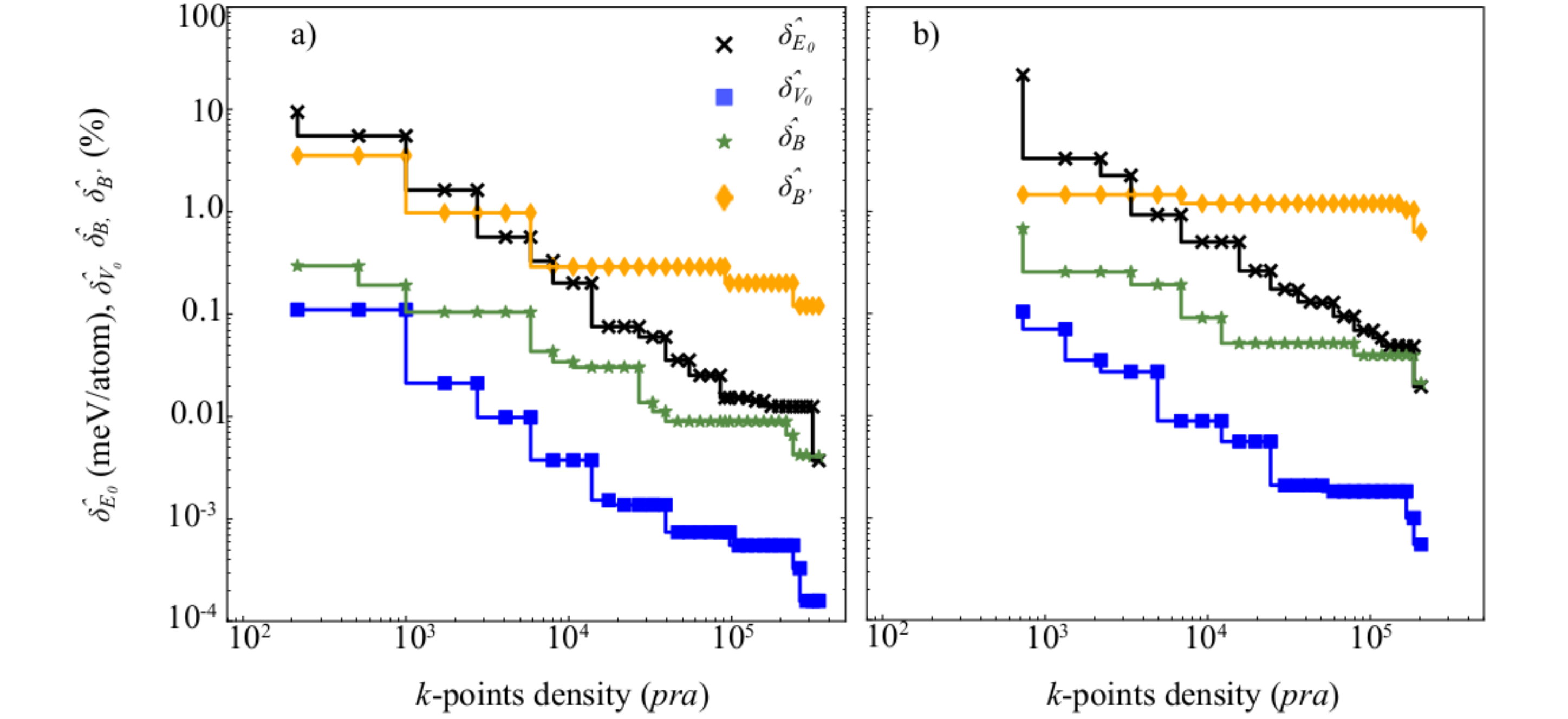}
  \caption{\label{fig:ParetoK} Pareto optimality fronts for the precision of the energy, $\hat{\delta}_{E_0}(k)$, volume, $\hat{\delta}_{V_0}(k)$, bulk modulus, $\hat{\delta}_{B}(k)$, and its pressure derivative, $\hat{\delta}_{B'}(k)$ in (a) VASP and (b) DMol3 for bcc Cr, as defined by Eq.~\eqref{eq:ParetoChoiceK}. These plots show that the maximum  precision remains constant for a series of $k$-point density choices and changes in a step-wise monotonic fashion that is similar for the both codes.}
\end{figure*}

\subsection{Convergence of properties with {\it k}-point density}
\label{Kpoints}

We now analyze how the  precision of the materials properties, $\delta_P$, converges with increasing $k$-point density. Since we are interested in changes of the magnitude of the  precision, and not the sign, we consider the absolute value of the  precision $|\delta_{P}(k)|$. We refer to $|\delta_{P}(k)|$ as $\delta_{P}(k)$, unless mentioned otherwise. 

Fig.~\eqref{fig:Complete_K_Convergence} shows that the absolute value of the  precision of each property for all metals decreases approximately following a power-law as a function of $k$-point density,
\begin{align}
  \label{eq:PowerKpoints}
  \delta_{P}(k) &= ck^{-m_P}.
\end{align}
%The dotted orange line in Fig.~\ref{fig:Complete_K_Convergence} shows the 90$^\mathrm{th}$ percentile values of the  precision as a function of the $k$-point density.
The characteristic exponents of the decay, $m_P$, correspond to the slopes shown in the log-log plots of Fig.~\ref{fig:Complete_K_Convergence} for the various properties and follow the trend of $m_{E_0} > m_{V_0} > m_{B} > m_{B'}$ for both VASP and DMol3 in the PBE functional. The larger exponent, $m_P$, implies a faster rate of convergence with respect to the $k$-point density. We believe that the different convergence rates among the properties are because $B$ and $B'$ are higher order derivatives of the energy, $E_0$, and hence converge more slowly. We confirm the same trends for the LDA exchange-correlation functional (see supplementary material). The distribution of the precision of the energy is comparable to the results by Morgan {\it et al.} for total energies of nine different metals \cite{Morgan2018} and we expect that a refined choice of $k$-point grids can give even better convergence rates.

The convergence rate of VASP and DMol3 is slightly different, which is likely due to differences in the basis sets; in VASP we use a plane wave basis with a fixed cutoff energy while in DMol3 we use local atomic orbitals with a fixed cutoff radius. We also observe that various materials display different convergence of the properties with respect to $k$-point density. This is consistent with some metals requiring larger $k$-point densities than others to be converged. For example, fcc Cu and Al require the highest $k$-point density to converge the elastic tensor \cite{MP_properties_deJong2015}.    

We now present a method to determine the $k$-point density required to ensure a desired  precision for each of the properties. We take the supremum function of the  precision $\delta_P(k)$, to find the minimum $k$-point density, $k_\mathrm{min}$, needed that ensures a desired  precision in the property, $P$,
\begin{eqnarray}
  \label{eq:ParetoChoiceK}
  \hat{\delta}_P(k_\mathrm{min}) = \sup_{k\ge k_\mathrm{min}} \delta_P(k).
\end{eqnarray}
Fig.~\ref{fig:ParetoK} illustrates for bcc Cr that $\hat{\delta}_P(k_\mathrm{min})$ is a monotonically step-wise decreasing function, forming a Pareto optimality front. We use this function to select the $k$-point density corresponding to the desired value of the  precision of the property. In other words, we have a new convergence guideline: increase the $k$-point density until the desired condition of  precision, say 1\%, is met for the property of interest. Furthermore, from each of the steps of this function, we can obtain the lowest possible $k$-point density, $k_\mathrm{min}$, for a desired  precision in the given property for the given material. The use of this lowest possible $k$-point density minimizes the computational cost to attain the desired  precision in the computed property.  

\begin{figure*}
  \includegraphics[width=0.65\textwidth]{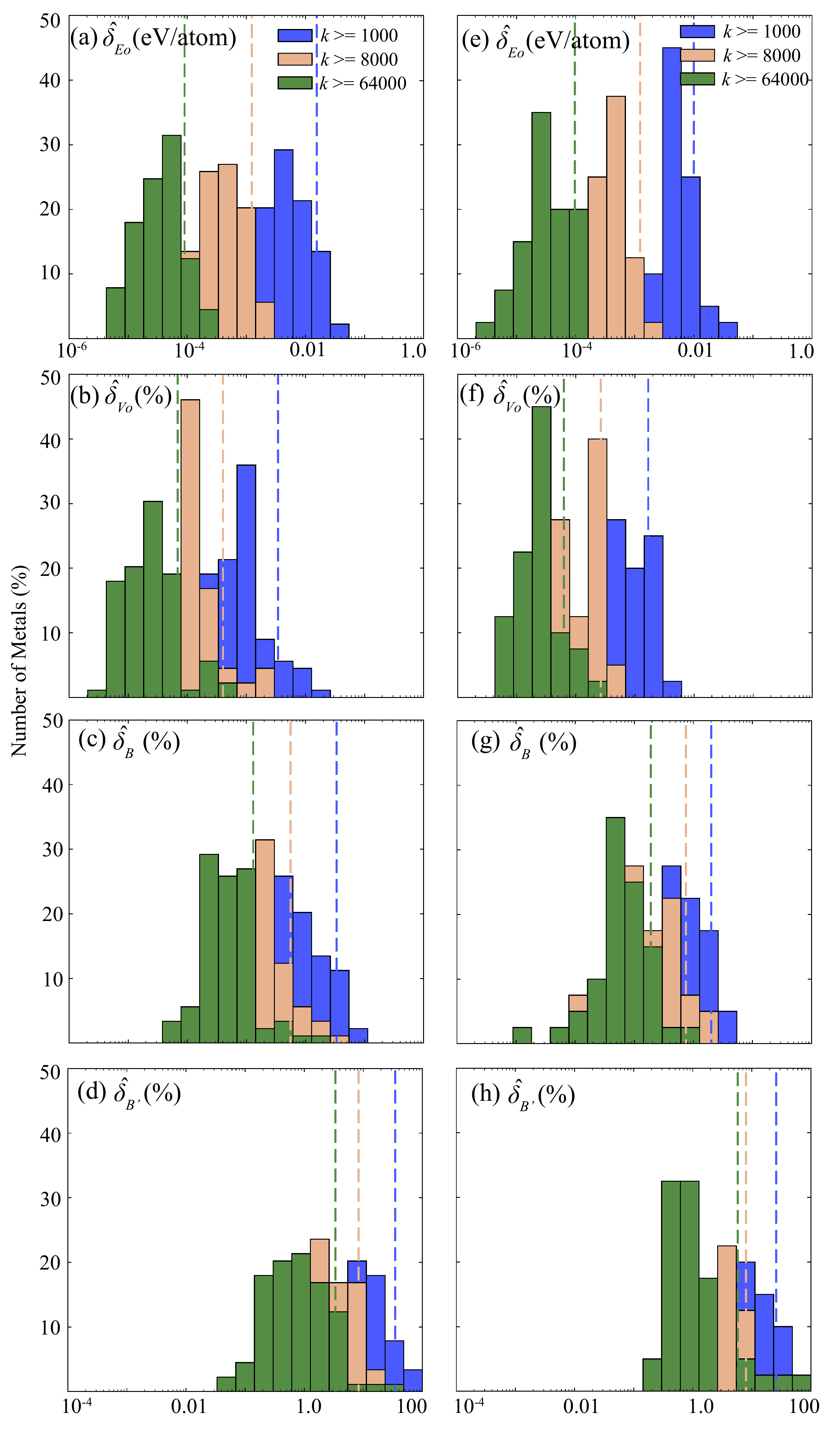}
  \caption{ \label{fig:Histograms_kpoints} The distribution of error (90$^\mathrm{th}$ percentile shown as dotted vertical lines) for the three choices of $k$-point density of 1,000 (blue), 8,000 (orange) and 64,000 (green) pra for $E_0$ calculated for PBE with (a) VASP and (e) DMol3. Similarly, the distributions for $V_0$ are shown in (b) and (f), for $B$ in (c) and (g), and for $B'$ in (d) and (h). A common choice of 1,000 to 10,000 $k$-points pra gives a precision of 1~mev/atom, 0.1\%, 1\% and 10\% for $E_0$, $V_0$, $B$, and $B'$ respectively. $B'$ requires more than $10^{4}$ $k$-points pra to converge to less than 1\%.}
\end{figure*}

\begin{figure*}[tbh]
  \includegraphics[width=0.65\textwidth]{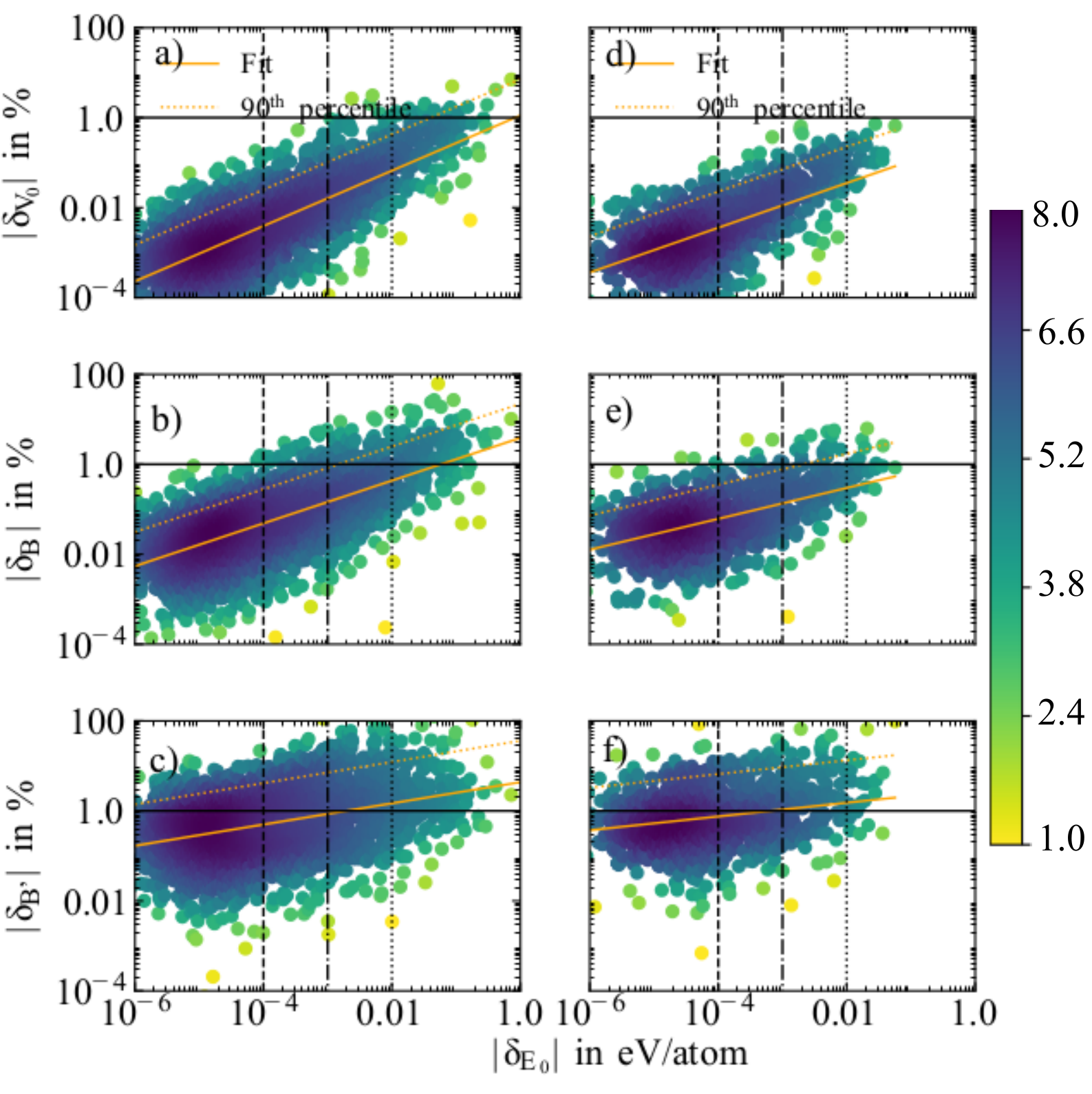}
  \caption{ \label{fig:Complete_E_Convergence} Correlation of precision error with energy convergence:
    The precision of $V_0$ as a function of the energy precision in eV/atom is shown as dots shaded from yellow to blue, calculated with the PBE functional using (a) VASP for 90 metals and (b) DMol3 for a subset of 40 metals. Similarly, the precision of $B$ is shown in (c) and (d) and of $B'$ in (e) and (f). The shading of the dots corresponds to the Gaussian kernel density estimate of the density of data points, yellow being one-eight the density as the dark blue dots. The solid orange lines show the fit of Eq.~\eqref{eq:PowerEnergy}. The 90$^\mathrm{th}$ percentile (dotted orange line) is a trend of the same slope as Eq.~\eqref{eq:PowerEnergy} drawn such that 90\% of the data points falls below the line. The vertical dotted lines indicate the three different choices for $\delta_{E_0}$ of 10, 1, and 0.1~meV/atom. For $\delta_{E_0}$=1~meV/atom, the precisions for $V_0$, $B$, and $B'$ are less than 0.1\%, 1.0\%, and 10.0\%, respectively, for 90\% of the metals. }
\end{figure*}

The distribution of the Pareto fronts given by Eq.~\eqref{eq:ParetoChoiceK} for all materials in our study provides insight into the precision that can be expected for a choice of $k$-points density. Fig.~\ref{fig:Histograms_kpoints} shows the distributions of this maximum precision error for three choices of the $k$-point density of 1,000, 8,000, and 64,000 pra. We note that the $k$-point density choices for the Materials Project, OQMD, and MaterialsWeb database are in the range of 1,000 to 10,000 pra. Fig.~\ref{fig:Histograms_kpoints} indicates that for a $k$-point density of 1,000 pra, 90\% of metals show precision errors of less than 1~meV/atom for $E_0$, less than 0.1\% for $V_0$, less than 1\% for $B$, and less than 10\% for $B'$. We note that in most of these databases higher $k$-point density choices are made for calculating the elastic constants especially for metals ($>$7,000 pra and up to 45,000 pra for metals like fcc Al and Cu in the Materials Project) for which we estimate precision errors of less than 1\% for all of $V_0$, $B$, and $B'$.

\begin{table}
\caption{\label{tab:precisions_kp} 90$^\mathrm{th}$ percentile of the distributions of the  precision of the energy, $E_0$, in meV/atom, and the equilibrium volume $V_0$, bulk modulus $B$ and its pressure derivative $B'$, in percent for the materials calculated with the PBE and LDA exchange-correlation functionals in VASP and DMol3 at the $k$-point density choices of 1,000, 8,000, and 64,000 pra.}
\begin{ruledtabular}
\begin{tabular}{ddddd}
\multicolumn{1}{l}{$k$-points} & \multicolumn{1}{c}{$\hat{\delta}_{E_0}$} & \multicolumn{1}{c}{$\hat{\delta}_{V_0}$} & \multicolumn{1}{c}{$\hat{\delta}_{B_0}$} & \multicolumn{1}{c}{$\hat{\delta}_{B'}$} \\
\colrule
\multicolumn{5}{c}{ -------- VASP with PBE -------- } \\
1000  & 15  & 0.3   & 3.0  & 30 \\
8000  & 1.0   & 0.03  & 0.5  & 9.0  \\
64000 & 0.09  & 0.005 & 0.1  & 3.0 \\
\colrule
\multicolumn{5}{c}{ -------- DMol3 with PBE -------- } \\
1000  & 10  & 0.2   & 2.0  & 20 \\
8000  & 1.0  & 0.03  & 0.5  & 8.0  \\
64000 & 0.09 & 0.005 & 0.1  & 6.0 \\
\colrule
\multicolumn{5}{c}{ -------- VASP with LDA -------- } \\
1000  & 20 & 0.1   & 1.0 & 10 \\
8000  & 1.0  & 0.01  & 0.2 & 3.0  \\
64000 & 0.1  & 0.005 & 0.1 & 1.0  \\
\colrule
\multicolumn{5}{c}{ -------- DMol3 with LDA -------- } \\
1000 & 10   & 0.1  & 1.0  & 4.0 \\
8000 & 1.0   & 0.02  & 0.2 & 1.0  \\
64000 & 0.1 & 0.008 & 0.07 & 0.5  \\
\end{tabular}
\end{ruledtabular}
\end{table}

\subsection{Convergence of properties with energy precision}
\label{Energy}

We now analyze how the precision of the structural and elastic properties correlates with the precision of the energy. DFT calculations of material properties often rely on the convergence of the energy to some predefined criteria such as 1~mRyd/atom or 1~meV/atom \cite{OnemevAccuracy1, OnemevAccuracy2}. However, it is difficult to know what energy convergence criterion is required to obtain the desired precision for other materials properties. We, therefore, determine how the precision of the energy correlates with the precision of the derived properties.

Fig.~\ref{fig:Complete_E_Convergence} shows that there is also an approximate power law decay for the convergence of the precision of the properties, $\delta_{P}(\delta_{E})$, as a function of the precision of the energy, $\delta_{E}$,
\begin{align} \label{eq:PowerEnergy}
\delta_{P}(\delta_{E}) &= c\delta_{E}^{-m_P}.
\end{align}
Compared to the correlation of the precision with the $k$-point density in Fig.~\ref{fig:Complete_K_Convergence}, the correlation of the property precision with the precision of the cohesive energy shows a larger spread. Nonetheless, the magnitude of the exponents, $m_P$, follows the same trend of $m_{V_0} > m_B > m_{B'}$ for both VASP and DMol3 in the PBE functional. We confirm this also for the LDA exchange-correlation functional (see supplementary material).

\begin{figure*}
  \includegraphics[width=0.85\textwidth]{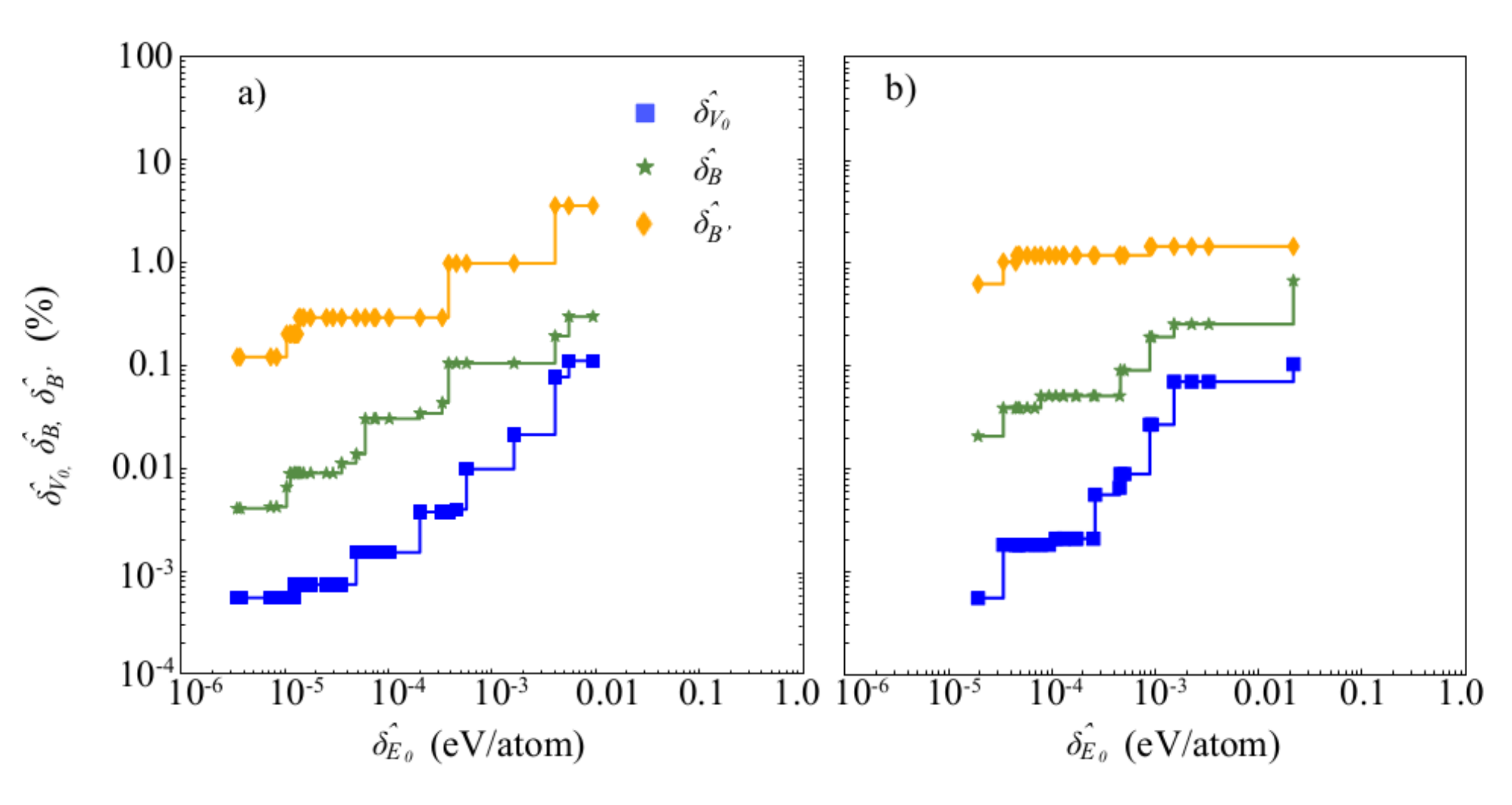}
  \caption{\label{fig:ParetoE} Pareto optimality front for bcc Cr of $\hat{\delta}_{E_0}(\delta_{E_\mathrm{max}})$, $\hat{\delta}_{V_0}(\delta_{E_\mathrm{max}})$, $\hat{\delta}_{B}(\delta_{E_\mathrm{max}})$, and $\hat{\delta}_{B'}(\delta_{E_\mathrm{max}})$ as a function of the energy precision $\delta_{E_0}(k)$ for (a) VASP and (b) DMol3 as defined by Eq.~\eqref{eq:ParetoChoiceE}}
\end{figure*}

Using the same approach as applied above to $\delta_P(k)$, we consider the supremum function of $\delta_P(\delta_E)$, to find the largest value of the energy precision, $\delta_{E_\mathrm{max}}$ that ensures a desired precision of the property, $P$,
\begin{eqnarray}
  \label{eq:ParetoChoiceE}
  \hat{\delta}_P(\delta_{E_\mathrm{max}}) = \sup_{\delta_E \le \delta_{E_\mathrm{max}}} \delta_P(\delta_E)
\end{eqnarray}
Fig.~\ref{fig:ParetoE} illustrates for bcc Cr that $\hat{\delta}_P(\delta_{E_\mathrm{max}})$ is again a monotonically step-wise increasing function, also forming a Pareto optimality front. We use this function to choose an energy convergence criterion corresponding to the desired value of the  precision of the property. In other words, we need to decrease the  precision error of the energy, until the desired precision, say{\it e.g.} 1\%, is met for the property of interest. Furthermore, from each of the steps of this function, we can obtain the largest value of the energy precision, $\delta_{E_\mathrm{max}}$ that provides the desired property precision for the given material. The use of this highest possible energy precision criterion is a means to benchmark the precision of the properties based on the known energy precision. This Pareto optimality analysis once again demonstrates that property convergence is achieved when the  precision requirement is met.

\begin{figure*}
  \includegraphics[width=0.65\textwidth]{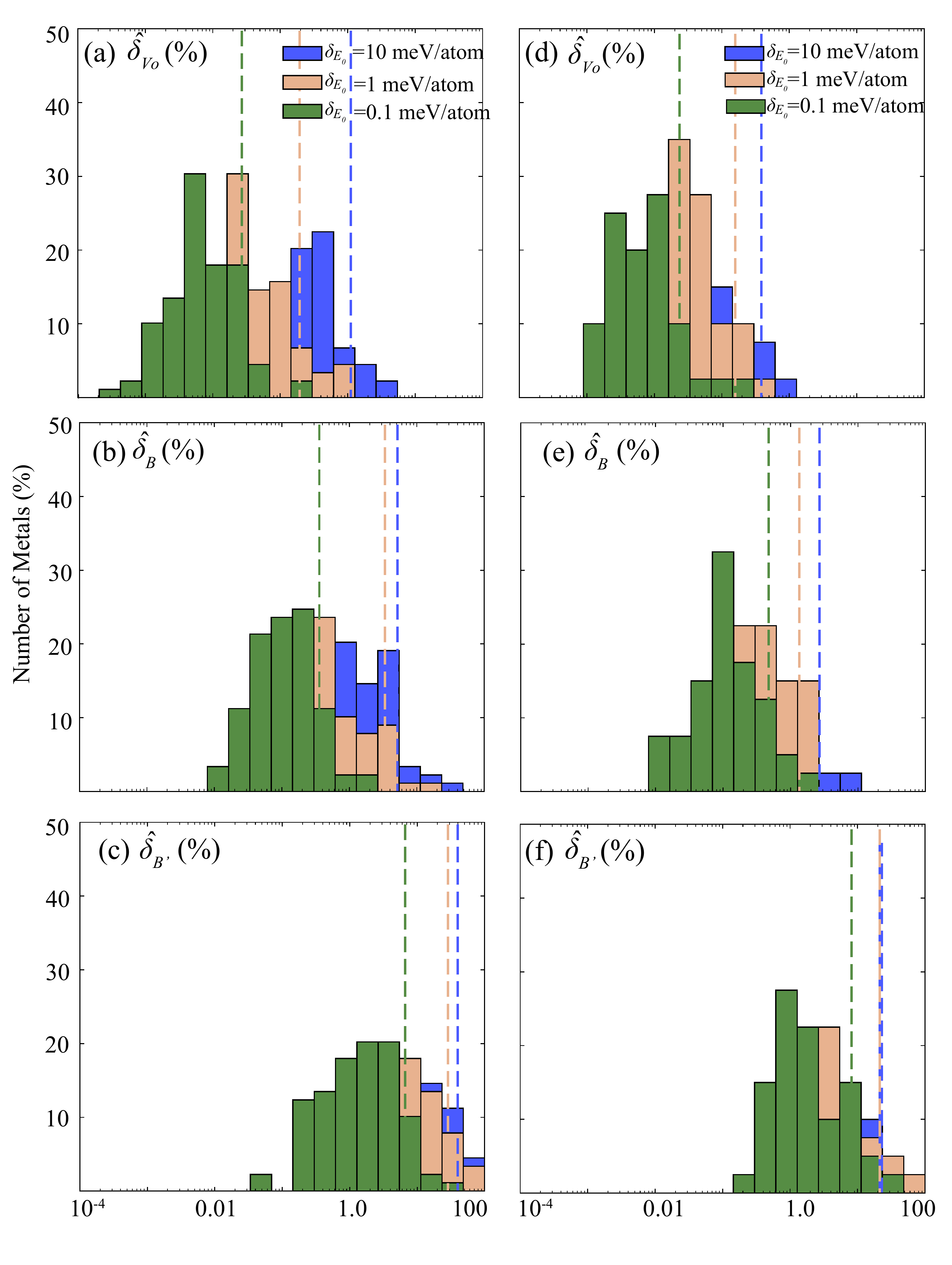}
  \caption{ \label{fig:Histograms_energy} The distribution of error (90$^\mathrm{th}$ percentile shown as dotted vertical lines) for 3 choices of energy precision of 10~meV/atom (blue), 1~meV/atom (orange) and 0.1~meV/atom (green) for $V_{0}$ calculated with (a) VASP at the PBE level and (d) DMol3 at the PBE level. Similarly $B$ is shown in (b) and (e) and $B'$ is shown in (c) and (f). A common choice of 1~meV/atom gives a  precision of 0.1 \%, 1 \% and 10 \% for $V_{0}$, $B$ and $B'$ respectively.  $B'$ requires a precision of 0.1~meV/atom to converge to less than 1\%.}
\end{figure*}

The histograms in Figs.~\ref{fig:Histograms_kpoints} and \ref{fig:Histograms_energy} show that the precisions for different $k$-point density choices and energy convergence criteria are approximately normal distributions for the set of metals calculated for VASP and DMol3. Tabs.~\ref{tab:precisions_kp} and \ref{tab:precisions_energy} compare the 90$^\mathrm{th}$ percentiles of the distribution for the precisions as a function of $k$-point density choice and energy precision, respectively, across the choice of code and exchange correlation functional.  We observe that the precisions of the energy, $E_0$, and volume, $V_0$, are very similar for both codes and exchange-correlation functionals. We do notice a slightly larger dependence on the code and exchange-correlation functional for $B$ and $B'$, especially when using the energy precision as convergence criteria. Among the properties, $B'$ requires the highest $k$-point density ($>$64,000 pra) or tighter energy convergence ($<$0.1 meV/atom) to attain a precision error of 1 \%, which is comparable to the precision error attained for $V_0$ and $B$ in materials databases.
%$B'$ changes by an order of magnitude whereas $V_0$ and $B$ change marginally for the tighter energy convergence criteria or the increased $k$-point density. \hl{I don't understand the last sentence. From Figure 6, it looks the other way around. The slopes for V0 and B are much steeper than for B'!}

We compare how the systematic decrease in the uncertainty of material properties $V_0$, $B$, and $B'$ for increasing $k$-point density correlates with the $k$-point density itself and with the convergence of the energy $E_0$. We find that the uncertainties in both cases follow approximately a power-law. We therefore recommend using the $k$-point density as the convergence parameter because it is computationally efficient and easy to handle as a direct input parameter in high-throughput frameworks for materials databases and because it correlates with precision at least as well as the energy.

\begin{table}
\caption{\label{tab:precisions_energy} 90$^\mathrm{th}$ percentile of the distributions of the  precision for the equilibrium volume, $V_0$, bulk modulus, $B$, and its pressure derivative, $B'$, in percent for the materials calculated with the PBE and LDA exchange-correlation functionals in VASP and DMol3 for energy convergence choices of $\hat{\delta}_{E_0}=$ 10, 1, and 0.1~meV/atom.}
\begin{ruledtabular}
\begin{tabular}{dddd}
\multicolumn{1}{c}{$\hat{\delta}_{E_0}$} & \multicolumn{1}{c}{$\hat{\delta}_{V_0}$} & \multicolumn{1}{c}{$\hat{\delta}_{B_0}$} & \multicolumn{1}{c}{$\hat{\delta}_{B'}$}\\
\colrule
\multicolumn{4}{c}{ -------- VASP with PBE -------- } \\
10  & 1.0   &  5.0  & 40 \\
1   & 0.1   &  3.0  & 30 \\
0.1   & 0.03  &  0.3  & 7.0 \\
\colrule
\multicolumn{4}{c}{ -------- DMol3 with PBE -------- } \\
10  &  0.5   &  3.0  & 20 \\
1   &  0.1   &  1.0  & 20  \\
0.1   &  0.02  &  0.4  & 8.0 \\
\colrule
\multicolumn{4}{c}{ -------- VASP with LDA -------- } \\
10  &  0.9    &  3.0  & 10.0 \\
1   &  0.03   &  1.0  & 8.0  \\
0.1   &  0.01   &  0.2  & 2.0  \\
\colrule
\multicolumn{4}{c}{ -------- DMol3 with LDA -------- } \\
10  &  0.2   &  3.0 & 8.0 \\
1   &  0.06  &  0.4 & 1.0  \\
0.1   &  0.02  &  0.1 & 0.8  \\
\end{tabular}
\end{ruledtabular}
\end{table}

\section{Conclusion}

In this article, we quantified the precision error of the cohesive energy, $E_0$, equilibrium volume, $V_0$, bulk modulus, $B$, and its pressure derivative, $B'$, for 29 transition metals and aluminum in three different crystal structures for density functional theory calculations. We found that the precision of these derived properties approximately correlates by a power law with both the $k$-point density and the precision of the cohesive energy. The rate of convergence of the properties follows the order $E_0 > V_0 > B > B'$. We showed that a Pareto optimality analysis of the precision provides a choice of a sufficient $k$-point density or energy convergence that ensures a desired precision in the derived properties. We predicted that a common choice of 8,000 $k$-points per reciprocal atom in high-throughput DFT databases provides a 90$^\mathrm{th}$ percentile precision of 1~meV/atom for the cohesive energy, 0.1\% for the volume, 1\% for the bulk modulus, and 10\% for the pressure derivative of the bulk modulus for the transition metals and aluminium. For the convergence of the property values with respect to the $k$-point density, we showed that energy convergence could provide a useful stopping criterion. We recommend the $k$-point density as the convergence parameter because it is computationally efficient, easy to use as a direct input parameter in high-throughput frameworks for materials databases, and correlates with precision at least as well as the energy. We also showed that these trends are not strongly dependent on the DFT code or exchange-correlation functional. We expect that the quantified uncertainties will help guide the determination of materials trends and the selection of materials.

%---------------------------------------------------------------------
\section{Data Availability}

The data required to reproduce these findings are available to download from Mendeley Data at  \url{https://doi.org/10.17632/p7dt4bjjmd.1}.

% ---------------------------------------------------------------------
\section{Acknowledgments}

This work was supported by the National Institute of Standards and Technology (NIST) under award 00095176 and by the National Science Foundation under grants Nos.\ DMR-1748464 and OAC-1740251. This research used computational resources provided by the University of Florida Research Computing (\url{http://researchcomputing.ufl.edu}) and the Texas Advanced Computing Center under Contract TG-DMR050028N.  This work used the Extreme Science and Engineering Discovery Environment (XSEDE), which is supported by National Science Foundation grant number ACI-1053575.

%The raw data is available at \hypertarget{NIST Research data}{http://researchdata.nist.gov/Materials-/Numerical-Precisions-of-Material-Properties-from-DFT}. \hl{This website is not working, we should leave it out and can add it if it is working by the time the paper gets published. Alternatively, we can upload the data to MaterialsWeb or MendeleyData.}  \hl{idea for graphical abstract, figure I presented at TMS for the intro} 

%%%%%%%%%%%%%%%%%%%%%%%%%%%%%%%%%%%%%%%%%%%%%%
% Bibliography
%%%%%%%%%%%%%%%%%%%%%%%%%%%%%%%%%%%%%%%%%%%%%%
\bibliography{biblio}

\end{document}